\documentclass{elsarticle}

\usepackage{latexsym}
\usepackage{amsmath}
\usepackage{amsfonts}
\usepackage{amssymb}
\usepackage{graphicx}
\usepackage[latin1]{inputenc}
\usepackage{color}

\begin{document}

\parindent 0mm
\parskip   2mm
\renewcommand{\baselinestretch}{1.0}


\date{}


\title{Nonlinear Economic State Equilibria via van der Waals Modeling} 


\author[epfl]{Max-Olivier Hongler\corref{cor}}
\ead{max.hongler@epfl.ch}
\address[epfl]{Ecole Polytechnique F\'ed\'erale de Lausanne (EPFL), School of Engineering (STI), Microengineering Institute, Station 17, CH-1015 Lausanne, Switzerland}
\author[unil]{Olivier Gallay\corref{}}
\ead{olivier.gallay@unil.ch}
\address[unil]{University of Lausanne, Faculty of Business and Economics (HEC Lausanne), Department of Operations, Quartier UNIL-Chamberonne, CH-1015 Lausanne, Switzerland}
\author[salus]{Fariba Hashemi\corref{}}
\ead{fariba.hashemi@outlook.com}
\address[salus]{Salus Partners, Avenue du Grey 117, CH-1018 Lausanne, Switzerland}
\cortext[cor]{Corresponding author}


\begin{abstract}

\noindent The renowned van der Waals (VDW) state equation quantifies the equilibrium relationship between  pressure $P$, volume $V$ and temperature $k_{B}T$ of a real gas.  We assign new variable interpretations adapted to the economic context: $P \rightarrow Y$, representing price; $V \rightarrow X$, representing demand; and $k_{B}T \rightarrow \kappa$, representing income, to describe an economic state equilibrium. With this reinterpretation, the price elasticity of demand (PED) and the income elasticity of demand (YED) are non-constant factors and may  exhibit a singularity of the cusp-catastrophe type. Within this economic framework, the counterpart of VDW liquid-gas phase transition illustrates a substitution mechanism  where one product or service is replaced by an alternative substitute. The  conceptual relevance of this reinterpretation  is discussed qualitatively and quantitatively  via several illustrations ranging from transport (carpooling), medical context (generic versus original medication) and empirical data drawn from the electricity market in Germany.
\end{abstract}


\maketitle
\thispagestyle{empty}
\vspace*{-2mm}


\section*{Prologue}\label{PROLOGUE}

\noindent  \textit{In 2023, the van der Waals (VDW) equation, a cornerstone of thermodynamics, celebrated its 150$^{\,th}$ anniversary. Originating from the doctoral thesis of the Nobel laureate Johannes Diderik van der Waals in 1873, this equation extends the ideal gas equation ($PV = {\cal R}T$) and offers profound insights into gas-liquid phase transitions and the underlying interplay of molecular forces. For an ideal gas at equilibrium, a decrease in volume corresponds to an increase in pressure, akin to dynamics observed in markets where price increases tend to dampen demand. As a modest tribute at the 150$^{\,th}$ anniversary occasion, we tentatively extend this economic analogy to real gas state equilibrium. To our surprise, the relevance of this a priori incongruous analogy has also been recently discussed by the late French astrophysicist Fran\c{c}ois Roddier (1936-2023) in a paper entitled "L'\'{e}quation de van der Waals appliqu\'{e}e \`{a} l'\'{e}conomie", \cite{roddier2017}.}


\section{Introduction}\label{INTRO}

\noindent  At thermostatic equilibrium, {\it "the absolute pressure $P$ exerted by a given mass of an ideal gas is inversely proportional to the volume $V$  it occupies if the temperature and amount of gas remain unchanged within a closed system"}, \cite{levine1978}. Formally  this ideal gas equilibrium (IGE)  is encapsulated in the {\it state equation} ${\cal R}_{g}(V, P)$:

\begin{equation}
\label{PVK}
{\cal R}_{g}(V, P):= P V = \kappa \qquad \qquad ({\rm Boyle-Marriott  \,\, law}),
\end{equation}

\noindent  where $\kappa \in \mathbb{R}^{+}$ is a proportionality constant.  Eq.(\ref{PVK}) quantifies  the  interplay between the macro-variables $P$ and  $V$ and the constant $\kappa$  is   proportional to the gas  temperature $T$ at the thermostatic  equilibrium.  We observe that in Eq.(\ref{PVK}),  the $P$ and $V$ variables play a perfectly symmetric role. {\it Ceteris paribus},  one may  view $P$ as  a $V$-dependent variable, namely $P\equiv P(V)$, or conversely one may  write $V \equiv V(P)$. For a given $\kappa$ (or equivalently $T$),  the relation ${\cal R}_{g}(V, P)= \kappa$ defines  an {\it isotherme}  line lying in  the positive quadrant of the real  plane $\mathbb{R}^{2}$.

\vspace{0.3cm}
\noindent  The thermostatic equilibrium given by Eq.(\ref{PVK})  bears  more than a passing resemblance to market  equilibrium (i.e., {\it market clearance}) which correlates   price $Y\in \mathbb{R}^{+}$ with  demand $X\in \mathbb{R}^{+}$. Since the ideal gas equilibrium theory stated in Eq.(\ref{PVK}) has been very successively extended to the van der Waals (VDW) real gases, it looks natural to infer the  potential relevance of the VDW equation in the economic context. This analogy already motivated several previous discussions \cite{gumjudpai2018, jammernegg1986, rashkovskiy2021, saslow1999}, and more particularly Fran\c{c}ois Roddier's recent work \cite{roddier2017}, which is closely related to the present paper. However, while in \cite{roddier2017} the focus is on macro-economic aspects, here we discuss the micro-economic implications.

Depending on whether we consider the dependent or independent variables,  we have dual interpretations:
\begin{itemize}
  \item[\textit{(i)}]  {\it when there is a price increase, equilibrium is maintained through a decrease in demand}, and vice versa. This is exemplified by \textit{sales}, which help to sustain demand during price shifts. 
  \item[\textit{(ii)}] {\it  when there is an increase in demand, equilibrium is maintained by an increase in price}, and vice versa. This is exemplified by the \textit{scarcity of resources}, which triggers competitive bidding.
\end{itemize}

\noindent Mirroring the gas equilibrium of Eq.(\ref{PVK}), the  market equilibrium is assumed to follow a state relation  ${\cal R}_{m}(X,Y)= \kappa$. For fixed  $\kappa$,  the state equation ${\cal R}_{m}(X,Y)= \kappa$ defines  {\it iso-$\kappa$} lines, and in the current context these lines are situated in the positive quadrant of the Cartesian plane $\mathbb{R}^{2}$.

\vspace{0.3cm}
\noindent A couple of basic, relevant measures in economics are the \textit{elasticity} factors:

\begin{equation}
\label{EDX} 
\left\{
\begin{array}{l}
{\cal E}_{Y/X} := \frac{\left\{ d \ln Y\right\}}{\left\{ d \ln X\right\}}  \\\\ 
{\cal E}_{X/Y} := \frac{\left\{ d \ln X\right\}}{\left\{ d \ln Y\right\}} = \frac{1}{{\cal E}_{Y/X}}\\\\

\end{array}
\right.
\end{equation}

\noindent The {\it elasticity } ${\cal E}_{Y/X}$ is hence a {\it sensitivity factor} which quantifies the {\it relative variation} of $Y$  in response to a  {\it relative variation} of $X$, and vice versa for ${\cal E}_{X/Y}$. 

\noindent For example, let us consider constant elasticity and the state equilibrium relation  ${\cal R}_{m}(X,Y)= X^{\alpha} Y^{\beta}$, with  $(\alpha, \beta)$ being two positive constants. In this case, for a fixed iso-$\kappa$ line (i.e., $d \kappa =0$),  the resulting  elasticity reads:

\begin{equation}
\label{EXPX}
\left\{
\begin{array}{l}
d\ln [{\cal R}_{m}(X,Y) ] = d \ln(\kappa) =0 \,\,\,  \Rightarrow  \,\,\,  \alpha \left\{d \ln X\right\} + \beta \left\{d \ln Y\right\} = 0\,\,\ \\\\   {\cal E}_{Y/X}=  \frac{\left\{d\ln Y\right\}}{\left\{d \ln X\right\}}  = - \frac{\beta}{\alpha} = {\rm constant}.
\end{array}
\right.
\end{equation}

\noindent Note that  by a suitable rescaling of  $X$ and $Y$, without loss of generality, we can express ${\cal R}_{m} (X,Y) = XY$ and therefore  $ {\cal E}_{Y/X} =  - 1$, which is the standard behavior  in  market equilibrium. For limited variation ranges of  $(X, Y)$, the  constant elasticity behavior given by Eq.(\ref{EXPX}) is perfectly appropriate. However, for extended parameter ranges, it   becomes imperative  to generalize  Eq.(\ref{EXPX}) and  allow   elasticity to become  state-dependent; this is developed in Section \ref{SECTHREE}. In Section \ref{SECFOUR}, economic variable interpretations, i.e., \textit{price elasticity of the demand} (PED) and \textit{demand elasticity of the income} (YED), is exposed. Several illustrations are provided in Section \ref{illustrations}. Concluding remarks are given in Section \ref{conclusion}.  
 \section{Non-constant elasticity - general framework}\label{SECTHREE}
 
 \noindent Let us here free ourselves from the sole market interpretation of the $(X,Y)$ variables and   develop a purely algebraic description directly inspired from the VDW thermostatic theory. Accordingly, we focus on state equations of the types:
   
\begin{equation}
\label{GLOBAL}
\left\{
\begin{array}{l}
(i)\quad  {\cal R}_{m}(X,Y) = XY  +  {\cal N}_{(X,\mu)}(X)= \kappa , \\\\
(ii)  \quad  {\cal R}_{m}(Y,X) = XY  +  {\cal N}_{(Y, \mu)}(Y)= \kappa,
 \end{array}
 \right.
 \end{equation}
 
 \noindent  where  $\mu$ stands for one (or possibly a set) of exogenous  control parameter(s). The extra nonlinear contributions  ${\cal N}_{(X,\mu)}(X)$ and  ${\cal N}_{(Y,\mu)}(Y)$ are introduced to   model  non-constant elasticity  responses.\\
 
Taking inspiration from the VDW generalization of the ideal gas equation, namely
 
 \begin{equation}
\label{VDWAA}
 ( {\rm ideal\,\, gases}) \Rightarrow PV = {\cal R}_{B} T \quad \longmapsto \qquad    ({\rm real \,\, gases})\Rightarrow \left[P + \frac{a}{V^{2}} \right] (V-b)  = {\cal R}_{B} T,
\end{equation}
 
\noindent let us explore the modelling relevance  offered by the class of state equations
 
\begin{equation}
\label{QUADRA12X}
\left\{
\begin{array}{l}
(i) \qquad {\cal N}_{(X, \mu)}(X) := \frac{a X}{ (X+ b)^{2}},  \quad  \\\\
(ii) \qquad {\cal N}_{(Y, \mu)}(X) := \frac{a Y}{ (Y+ b)^{2}}, 
\end{array} 
\right.
\end{equation}

\noindent where $  \mu := \left\{ a, b\right\}$, $a \in \mathbb{R}^{+}$, and  $b \in \mathbb{R}^{+}$.\\

With the specific choice in Eq.(\ref{QUADRA12X}),  Eq.(\ref{GLOBAL}) reads:

\begin{equation}
\label{QUADRA2X}
\left\{
\begin{array}{l}

(i) \, \, Y = Y(X) =\frac{\kappa}{X} - \frac{a}{(X+b)^{2}} = \,\,\, \Leftrightarrow \,\,\, {\cal R}_m(X,Y) := \left[ Y + \frac{a}{(X+b)^{2}} \right] X = \kappa, \\\\

(ii) \, X = X(Y) =\frac{\kappa}{Y} - \frac{a}{(Y+b)^{2}} = \,\,\, \Leftrightarrow \,\,\, {\cal R}_m(Y,X) := \left[ X + \frac{a}{(Y+b)^{2}} \right] Y= \kappa.
\end{array}
\right.
\end{equation}
 
 \noindent Eqs.(\ref{QUADRA12X}) and (\ref{QUADRA2X}) follow  straightforwardly from Eq.(\ref{VDWAA}) with the identifications $V \mapsto X+b$, $P \mapsto Y$ and ${\cal R}_B T \mapsto \kappa$. In particular, in Eq.(\ref{QUADRA2X}), the parameter  $a$ quantifies   the degree of inelasticity. For the iso-$\kappa$ line  to remain in the positive quadrant, we further impose $\kappa >\frac{a}{4b} $; the explicit derivation of this lower bound is postponed to the proof of Lemma 1 below.
 \noindent  Due to the  symmetric roles played by  $X$ and $Y$, we limit the analytical discussion to case \textit{(i)} (the dual case \textit{(ii)} follows identically from a $Y_r \leftrightarrow X_r$ substitution). Let us now introduce the rescaling $(X,Y, \kappa) \mapsto (X_r,Y_r, \kappa_r)$ defined as:

\begin{equation}
\label{REDUCEDX}
\left\{
\begin{array}{l}
Y_r : = \frac{Y}{Y_c}, \quad  X_r : = \frac{X}{X_c} \quad {\rm and} \quad \kappa_r : = \frac{\kappa}{\kappa_c}, \\\\
X_c = 2b, \quad Y_c= \frac{a}{27b^{2}} \quad {\rm and } \quad \kappa_c = \frac{8a}{27b}.
\end{array}
\right.
\end{equation}

\noindent  Standard algebra shows that when plugging  Eq.(\ref{REDUCEDX})  into  Eq.(\ref{QUADRA2X}) ), we obtain a form that is independent of any parameters:

\begin{equation}
\label{REDUCED1X}
 {\cal R}(X_r, Y_r)  :=  \left[ Y_r + \frac{27}{(2 X_r +1)^{2}}\right]  X_r= 4 \kappa_r \quad \Leftrightarrow \quad  Y_r  = \frac{4\kappa_r}{X_r} - \frac{27}{ \left[ 2X_r +1 \right]^{2} } .
\end{equation}

\begin{figure}[htbp] \label{WCUR22}
 \begin{center}
    \includegraphics[height=7cm]{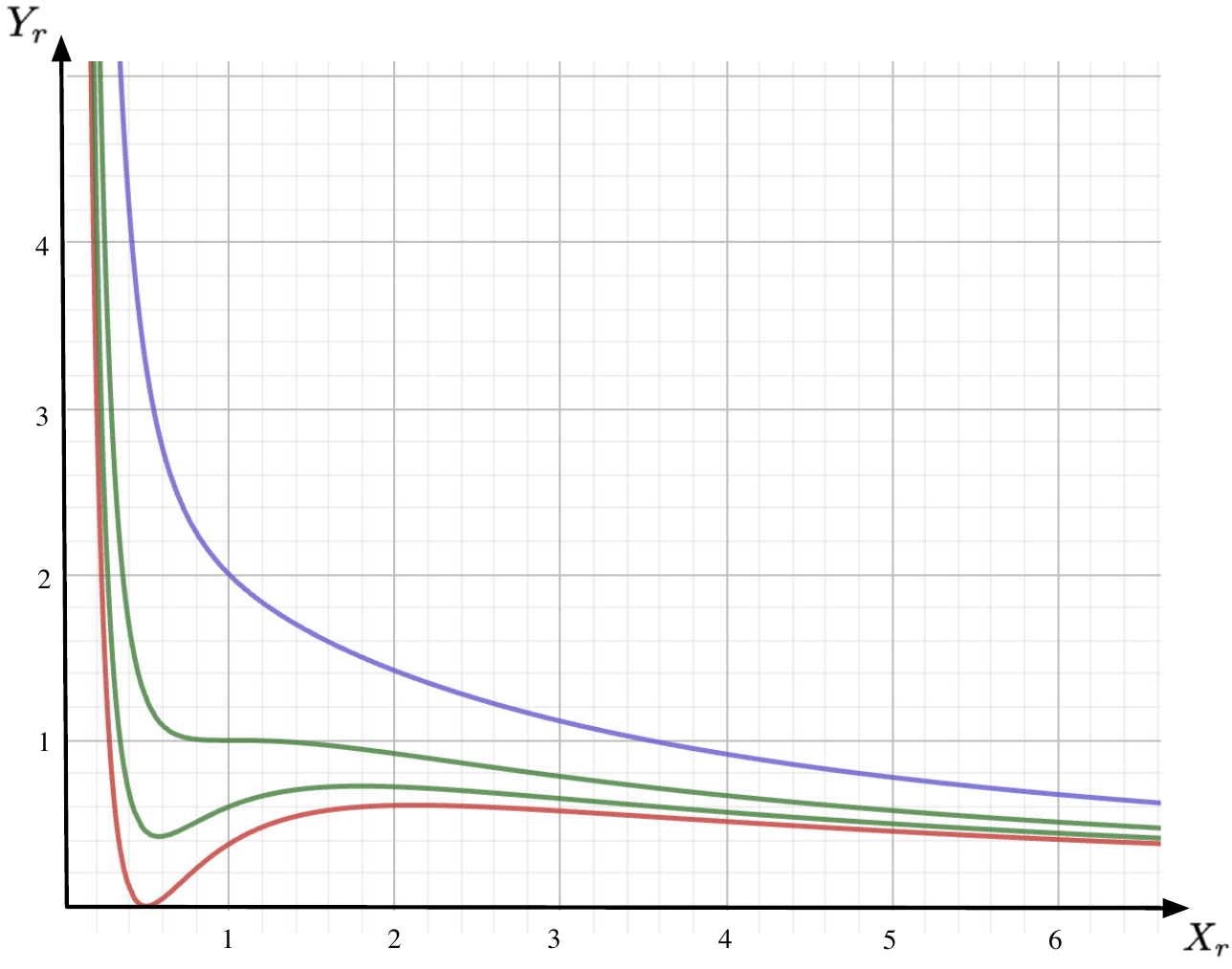}
    \caption{\scriptsize {Sketch of iso-$\kappa_r$ lines $Y_r$ (ordinate) as a function of $X_r$ (abscissa), as given by Eq.(\ref{REDUCED1X}). Here the iso-$\kappa_{r}$ lines are sketched for $\kappa_r =\frac{27}{32}$ (red), $\kappa_r=0.9$ (green, below), $\kappa_r =1.0$ (green, above, in this case the iso-$\kappa_r$ line exhibits a triple point at $X_r=1$), and $\kappa_r =1.1$ (blue). 
        }}
     \label{WC}
     \end{center}
 \end{figure}

\noindent  A selection of the iso-$\kappa_{r}$ lines obtained from Eq.(\ref{REDUCED1X}) are drawn in Figure \ref{WC}. Like for the VDW equation, the following features can be highlighted.

\newpage

\noindent {\bf Lemma 1}. {\it Eq.(\ref{REDUCED1X}) possesses the following properties:}

\begin{itemize}
  \item[\textit{(i)}]  {\it on the iso-$\kappa_r$ line  with  $\kappa_r=1$, the position  $(X_r, Y_r) =(1,1)$  defines an equilibrium {\it  triple  point} characterised by  $\frac{dY_r}{dX_r} =  \frac{d^{2}Y_r}{dX_r^{2}} \mid_{X_r=1} =0$.}
  \item[\textit{(ii)}] {\it for $\kappa_r > \frac{27}{32} $, all   iso-$\kappa_r$  lines remain in  the positive quadrant.}

\end{itemize}

\vspace{0.3cm}
\noindent{\bf Proof of Lemma 1}. 

\noindent \textit{(i)} is immediately verified by direct calculation. For \textit{(ii)}, one verifies  that for $\kappa = \frac{27}{32}$, we have:

$$
\left\{
\begin{array}{l}
Y_r\left(\frac{1}{2}\right) = \frac{dY_r}{dX_r}\mid_{X_r=\frac{1}{2}} =0,\\\\

\frac{d^{2}Y_r}{dX^{2}_r}\mid_{X_r=\frac{1}{2}} = 14.75>0,
\end{array}
\right.
$$

\noindent which shows that at  $X_r= \frac{1}{2}$, the iso-$\kappa_r$ line reaches its minimum and   $Y_r(\frac{1}{2}) = 0$. For all $\kappa_r > \frac{27}{32}$, the  minimum of the corresponding iso-$\kappa_r$ line is positive, implying the iso-$\kappa_r$ line to remain in the positive quadrant. This in turn implies that $\kappa_r > \frac{27}{32} \,\, \Rightarrow \,\, \kappa > \frac{a}{4b} $, as already  expressed in Eq.(\ref{QUADRA2X}).

\rightline{$\Box$}

\vspace{0.3cm}
\noindent {\bf Proposition 1}: ({\it ${\cal E}_{Y_r/X_r}$ and ${\cal E}_{X_r/\kappa_r}$  elasticities}).\\

\begin{figure}[htbp] \label{ELASTO}
 \begin{center}
    \includegraphics[height=8cm]{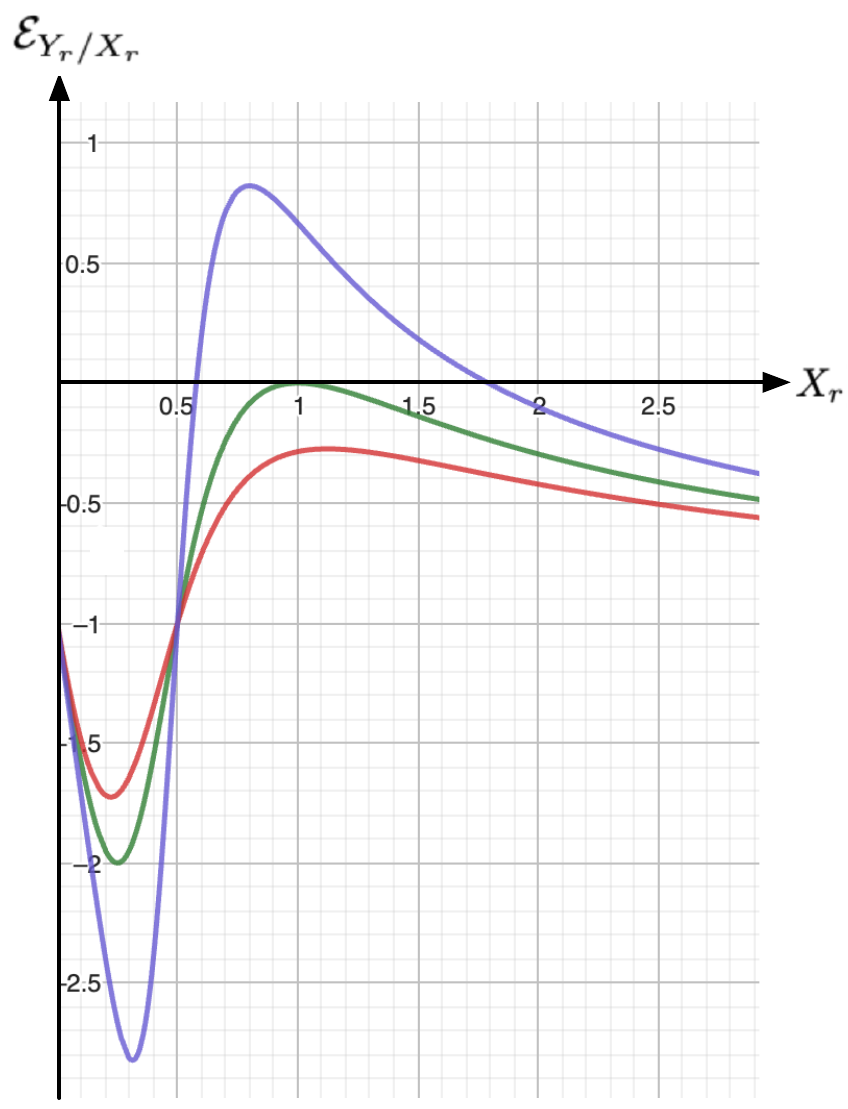}
    \caption{\scriptsize {Sketch of the elasticity  ${\cal E}_{Y_r/X_r}$ according to Eq.(\ref{REDUCED2X})  for  $\kappa_r = 0.9$ (red),  $\kappa_r = 1.0$ (green) and $\kappa_r = 1.1$ (orange). For $\kappa_r <1$, we systematically have ${\cal E}_{Y_r/X_r}>0$. According to Lemma 2 (see Appendix), we have ${\cal E}_{Y_r/X_r = \frac{1}{2}}=-1$ and ${\cal E}_{Y_r/X_r}<0$
   for $X_r \in \left[0, \frac{1}{2} \right]$.}}
     \label{ELASTOFIG}
     \end{center}
 \end{figure}

\noindent {\it For the state equilibrium given by Eq.(\ref{REDUCED1X}), and with the notation $Z(X_r) := \frac{1}{(2X_r +1)}$, we have:

\vspace{0.3cm}

\noindent (i)  on  iso-$\kappa_r$ lines, the corresponding elasticity reads (a selection of the elasticies obtained from Eq.(\ref{REDUCED2X}) are drawn in Figure \ref{ELASTO}):}

\begin{equation}
\label{REDUCED2X}
\begin{array}{l} 
 {\cal E}_{Y_r/X_r} :=  \frac{\left\{d \ln Y_r\right\}}{\left\{d \ln X_r\right\}} =  -  \left[ 
 \frac{
 \frac{4\kappa_r}{X_r} - 108 X_r [Z(X_r) ]^{3}}
 {
 {\frac{4\kappa_r}{X_r} - 27[Z(X_r) ]^{2}
}} \right] ,

 \end{array}
\end{equation}

\vspace{0.3cm}

\noindent {\it (ii) on iso-$Y_r$ lines,   the corresponding  elasticity    reads:}

\begin{equation}
\label{KELAS}
{\cal E}_{X_r /\kappa_r} := \frac{\left\{ d \ln X_r \right\}}{\left\{d \ln \kappa_r\right\}} =  \frac{1}{
1- \frac{108 X_r[Z(X_r)]^{3}}{Y_r + 27 [Z(X_r)]^{2}} 
}  .
\end{equation}

\vspace{0.3cm}
\noindent{\bf Proof of Proposition  1.}

\noindent {\it  (i) Fix an iso-$\kappa_r$ line and  take the logarithm of both sides of Eq.(\ref{REDUCED1X}). Then perform the corresponding variations, we have:

$$\ln\left[ Y_r \right] = \ln \left[ \frac{4\kappa_r}{X_r} - \frac{27}{(2X_r+1)^{2}}   \right]  \,\,\, \Rightarrow \,\,\, 
\frac{dY_r}{Y_r} = -   \left[ 
\frac{
 \frac{4\kappa_r}{X_r} - 
\frac{108\, X_r}{(2X_r +1)^{3}} }
{\frac{4\kappa_r}{X_r} - \frac{27}{(2X_r+1)^{2}} }
 \right]
 \frac{dX_r }{X_r},
$$

\noindent and, with the notation $Z(X_r) := \frac{1}{(2X_r +1)}$, the assertion  follows.

\vspace{0.3cm}
 \noindent  ii) Fix an iso-$Y_r$ line and  take the logarithm of both sides of Eq.(\ref{REDUCED1X}). Then perform the corresponding variations, we have:
 $$
\ln\left\{ \left[ Y_r + \frac{27}{(2 X_r +1)^{2}}\right]  X_r \right\} = \ln(4 \kappa_r) \,\,\, \Rightarrow \,\,\, \frac{dX_r}{X_r} \left[ 1 - \frac{108 X_r [Z(X_r)]^{3}}{Y_r + 27[Z(X_r)]^{2}}  \right] = \frac{d \kappa_r}{\kappa_r},
$$

\noindent and the assertion follows.
 }

\rightline{$\Box$}

\vspace{0.3cm}

\newpage
\noindent {\bf Proposition 2}: {\it (cusp catastrophe singularity).}

\noindent {\it  (i) Under the change of variables:}

\begin{equation}
\label{XX}
x :=\left[ \frac{2}{2X_r + 1} -\frac{2}{3}\right] \quad \Leftrightarrow \quad X_r =  \frac{3}{3x +2} - \frac{1}{2},
\end{equation}

\noindent  {\it Eq.(\ref{REDUCED1X}) reduces to  the cubic relation:}

\rightline{$\Box$}
\begin{equation}
\label{CUSP1X}
\left\{
\begin{array}{l}
x^{3} + u_2x + u_1 =0, \\\\
u_2 := \frac{4}{27}\left[ 8\kappa_r+ Y_r- 9 \right] \quad {\rm and } \quad u_1:= \frac{16}{81}\left[ 4\kappa_r -  Y_r-3\right] =0. 
\end{array}
\right.
\end{equation}

\vspace{0.3cm}

\begin{figure}[htbp]
 \begin{center}
    \includegraphics[width=13cm]{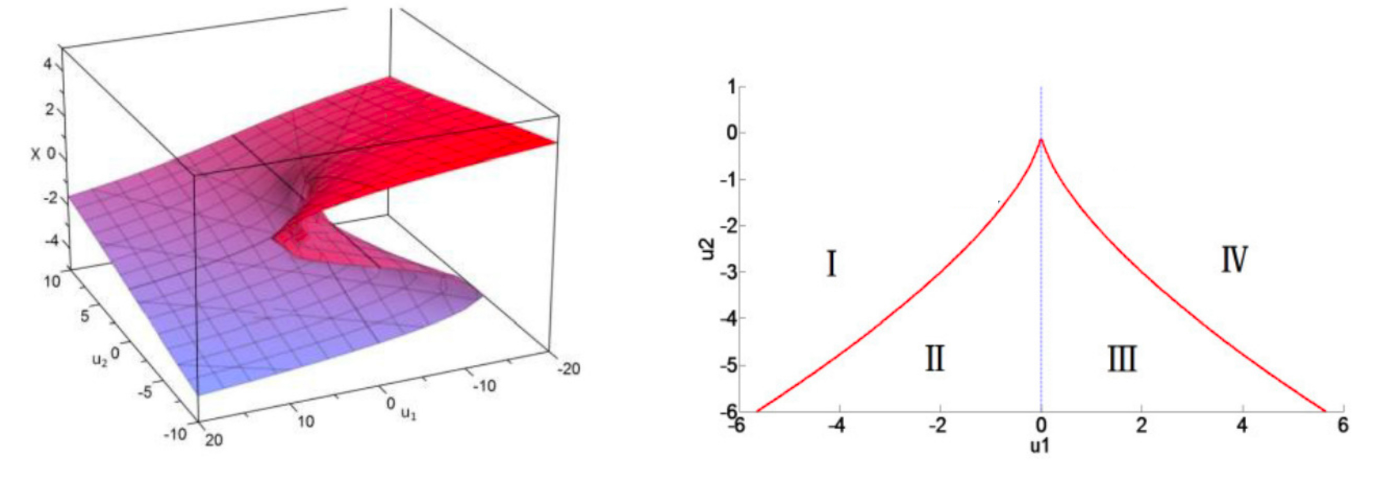}
    \caption{\scriptsize {On the left panel, \cite{huang2014}, the  position $(x,u_1, u_2)=(0,0,0)$ corresponds to  a triple point singularity. On the right panel, the red line is determined by  $\Delta =0$. This shows the projection of the sheet fold  on  the $(u_1, u_2)$ plane and it corresponds to a cusp catastrophe. For example, the couple of points $u_2 = -3$ and $u_1 = \pm 2$ lies on the red curve.   In the region $u_2 \in \mathbb{R}^{+} $,  we have $\Delta>0$ and here the relation ${\cal R}(X_r, Y_r)$ of Eq.(\ref{REDUCED1X}) is one-to-one  yielding  ${\cal E}_{Y_r/X_r} <0$. For $u_2<0$,   the relation ${\cal R}(X_r, Y_r)$  is surjective in  regions II and III where  $\Delta <0$ and   ${\cal E}_{Y_r/X_r} <0$. In regions I and IV,  we have $\Delta>0$ and ${\cal E}_{Y_r/X_r} <0$.
    }}
     \label{SHEE1}
     \end{center}
 \end{figure}

\noindent {\it  (ii) For $ \Delta:= - \left[ 4u_2^{3} + 27 u_1^{2}\right]$, the 
 cubic equation of Eq.(\ref{CUSP1X}) admits one, respectively three, real solutions. The limiting case  $u_1= u_2 =0 \,\, \Rightarrow \,\, \Delta =0$  describes a cusp catastrophe singularity as sketched in Figure \ref{SHEE1}.}

\vspace{0.3cm}
\noindent {\bf Proof of Proposition 2}.\\
\textit{(i)} is verified by direct calculation and \textit{(ii)} is the direct consequence of the Cardan classical theory for cubic equations.

\rightline{$\Box$}

\noindent An additional list of properties of Eq.(\ref{REDUCED1X}) can  be found in the Appendix. 


\section{Price elasticity of demand (PED) and income elasticity of demand (YED)} \label{SECFOUR}

\noindent  For a comprehensive description of the  economic equilibrium state equation  
given by Eq.(\ref{REDUCED1X}), with $X_r$ being the demand (or the inverse demand)and $Y_r$ being the price, it is mandatory to assign an ad-hoc meaning to  $\kappa_r$ (i.e., the temperature in the nominal VDW model). The yet missing variable in the economic  modeling context is a budget (or income) variable, and it is therefore natural to assign to $\kappa_r$  this complementary dimension. Summarizing, from section \ref{SECTHREE}, one concludes that the state  equilibrium binding the price $Y_r$, the demand $X_r$ and the budget $\kappa_r$ is geometrically encapsulated whiting a sheet ${\cal S}:= {\cal S}(X_r,Y_r,\kappa_r)$  immersed  within  the 3D space (see  Figure \ref{SHEE1}). According to  Eq.(\ref{CUSP1X}),  this  sheet possesses  a  fold singularity located at the  triple point $(X_r,Y_r,\kappa_r)= (1,1,1)$. From this description, a couple of elasticity factors can be naturally defined:

\begin{itemize} 
\item[\textit{(i)}] {\bf PED-elasticity}. For {\it iso-budget lines} (i.e., fixed  budget $\kappa_r$), the  PED-elasticity ${\cal E}_{Y_r/X_r}$ denotes the {\it price elasticity of demand}; it is given by Eq.(\ref{REDUCED2X}) and it is sketched in Figure \ref{ELASTO}.

\item[\textit{(ii)}] {\bf YED-elasticity}. For {\it iso-price lines} (i.e., fixed  price $Y_r$), the  YED-elasticity ${\cal E}_{X_r/\kappa_r}$  is  the  {\it income elasticity of demand};  it  is given by Eq.(\ref{KELAS}). As shown in Figure 4, the YED factor enables to distinguish between the {\it normal goods}, for which  ${\cal E}_{X_r/\kappa_r}>0$, and the {\it inferior goods}, which correspond to situations where ${\cal E}_{X_r/\kappa_r}<0$.


\end{itemize} 

\begin{figure}[htbp] \label{KELASFIG}
 \begin{center} \includegraphics[height=7cm]{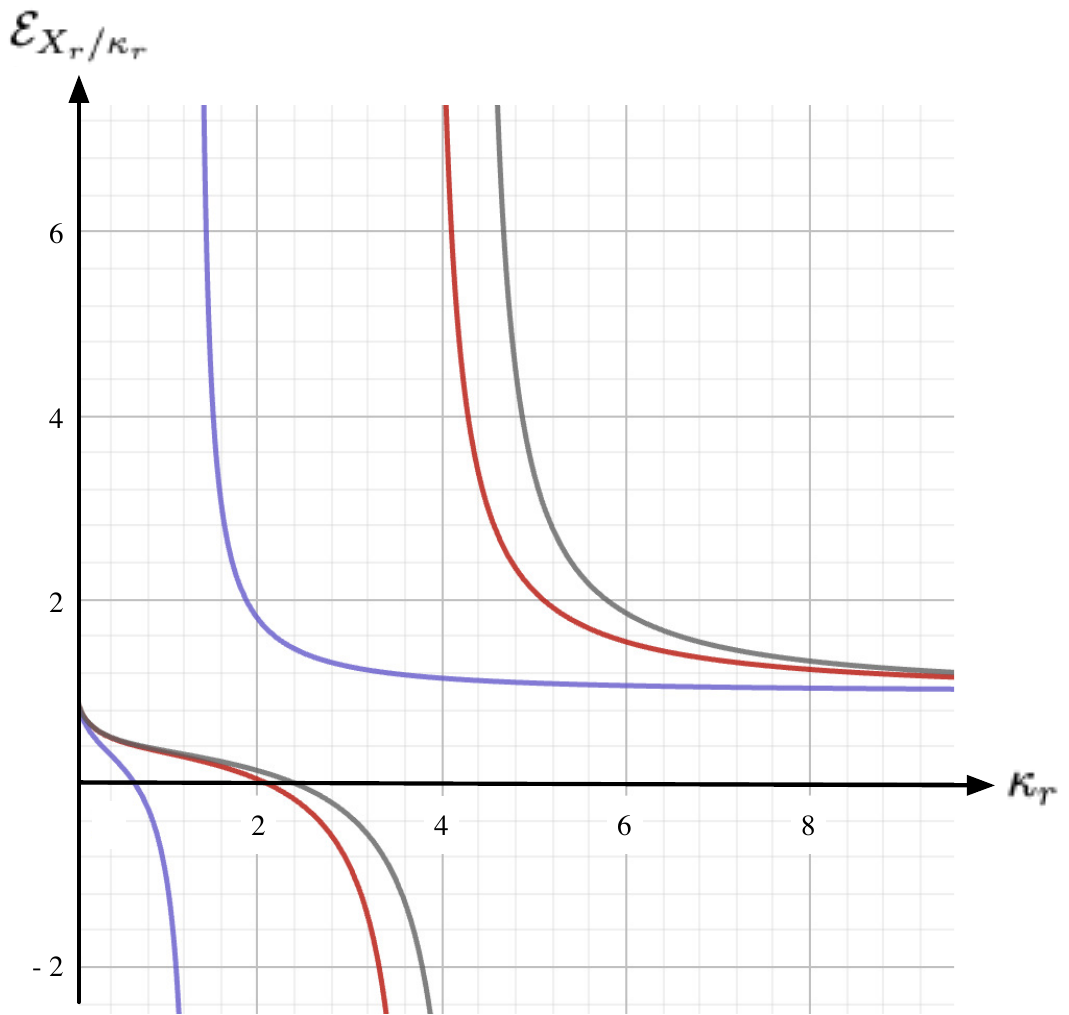}
    \caption{\scriptsize{The YED  ${\cal E}_{X_r/\kappa_r}$  (ordinate) as a function of $\kappa_r$ (abcissa), according to  Eq.(\ref{KELAS}), for $Y_r=0.8$ (grey), $Y_r=1.0$ (red) and $Y_r=5$ (blue). For large  price $Y_r$}, we have $\lim_{Y_r \rightarrow \infty} {\cal E}_{X_r/\kappa_r} = +1$, showing that at high price we systematically have a {\it normal good} behaviour, namely ${\cal E}_{X_r/\kappa_r}>0$; for intermediate $Y_r$, there exists a $Y_r$-dependent range where ${\cal E}_{X_r/\kappa_r} <0$, which is the signature of an {\it inferior good} behavior; for low $Y_r$, the normal behavior ${\cal E}_{X_r/\kappa_r}>0$ is recovered.}
     \end{center}
 \end{figure}

\subsection{Substitution phenomena and Maxwell-type plateau of prices} \label{SCARCITY}


\noindent Let us now focus on a usual economic state equilibrium where an increase in demand triggers an increase in prices. In this situation, the corresponding price elasticity of the demand will be positive definite (a practical illustration for this is the German electricity market, which is further discussed in Section \ref{ELECTRO}). To qualitatively recover the typical VDW behavior of Eq.(\ref{REDUCED1X}), let us interpret $X_r$ as the  inverse demand (i.e., rather than the demand itself) while keeping $Y_r$ to represent the price. With this inverse demand convention and since  $\ln[X_r] = - \ln[1/X_r]$, the corresponding price elasticity of the inverse demand becomes systematically negative for $\kappa_r >1$, as  sketched in Figure \ref{WC}\footnote{The state equilibrium description  in terms of the inverse demand rather than the demand itself is actually adopted in applications as in \cite{wan2022}.}. For $\kappa_r<1$ however, ${\cal E}_{Y_r/X_r}>0$  within a $\kappa_r$-dependent range of the inverse demand  $X_r$. This would hence describe a range of economic equilibriums in which increases of the demand generate increases of the price that do not reflect the actual reality. Inspired directly from the VDW liquid-gas phase transition, instead of following the $z$-shape iso-budget line, we describe the state equilibrium by constructing, for each iso-budget line with $\kappa_r<1$, a price plateau at a $\kappa_r$-constant level $Y_r:= y_{\kappa_r}$. In the VDW case, extra physical considerations fix the position of the plateau level $y_{\kappa_r}$ by using the reknowned Maxwell construction:

\begin{equation}
\label{MAXWELL}
\left\{
\begin{array}{l}
Y_r\left(l_{\kappa_r}\right) = Y_r\left(h_{\kappa_r}\right) = y_{\kappa_r}, \\\\
 l_{\kappa_r}\leq X_{r,{\rm min}} \quad {\rm and } \quad  h_{\kappa_r}\geq X_{r,{\rm max}}
\\\\
\int_{l_{\kappa_r}}^{h_{\kappa_r}} Y(X_r) dX_r = [h_{\kappa_r}-l_{\kappa_r}] y_{\kappa_r}
\end{array}
\right.
\end{equation}

\noindent where $l_{\kappa_r}\leq h_{\kappa_r}$   denotes the intersection abscissa of the $y_p$ horizontal line with the iso-budget line $\kappa_r<1$, and $X_{r,{\rm min}}$ and $X_{r,{\rm max}}$ are the  abscissa of extrema of the same $\kappa_r$ iso-budget line. While the Maxwell construction given in Eq.(\ref{MAXWELL}) also arises in economic contexts (see for example Figure 2 in \cite{guimaraes2011}, with  related explanations), a more rigorous justification yet remains open in general. The important feature is the necessity to construct a plateau for iso-budget lines with $\kappa_r <1$. This plateau behavior in the economic context effectively models a substitution mechanism during which a nominal product (or service) is, due to a  demand modification, progressively replaced by an available substitute. The actual position on the plateau itself informs on the relative  proportion of the nominal product with respect to its substitute. This stands in full analogy with the VDW context where the plateau describes the liquid and gas mixture of phases.

\section{Illustrations}\label{illustrations}

\noindent Far from exhaustive, yet aimed at sparking imagination, we now present a few examples where the VDW equation of state provides insightful modeling perspectives in an economic context.

\subsection{Transportation context - carpooling substitution}

\noindent In the context of transportation, specifically regarding a carpooling commuting policy, let's consider the variables as follows: $Y_r$ represents the transportation price, $X_r$ signifies the inverse transportation demand, and $\kappa_r$ denotes the per capita budget. Given these definitions, we can describe the qualitative stylized behavior as follows:

\begin{itemize}
  \item[\textit{(a)}] Wealthy commuters are characterized by $\kappa_r >1$. We assume that these individuals consistently choose to travel alone in their private cars. In this case, an increase in transport demand (which is represented by a decrease in $X_r$, the inverse transportation demand) leads to an equilibrium state through an increase in $Y_r$, the transportation price, as governed by Eq.(\ref{REDUCED1X}). In the context of the nominal VDW equation, this state of equilibrium can be analogized to a condition of high temperature. 

  \item[\textit{(b)}]  
For ordinary commuters, characterized by $\kappa_r < 1$, the high cost of transportation is a significant burden. These individuals are thus more inclined to embrace a carpooling policy as a more economical, but also environmentally friendly option. This shift towards carpooling helps stabilize both transportation prices and the ecological footprint, creating a scenario corresponding to the plateau phase in the VDW equation for real gases. Within the transportation context, a substitution is struck between individual commuting and carpooling. As the demand for transportation increases (indicated by a decrease in $X_r$), there is a corresponding rise in the proportion of commuters opting for carpooling. This dynamic is represented by a movement to the right on the plateau, where the transportation price $Y_r$ remains constant despite the changing demand. This analogy extends to the nominal VDW model for real gases, where the position on the VDW plateau indicates the ratio of liquid to gas in a mixture. Similarly, in the transportation model, the position on the plateau reflects the balance between individual commuting and carpooling, illustrating how economic and environmental considerations can lead to a stable equilibrium in commuter behaviors.
  
    \item[\textit{(c)}] When all opportunities for carpooling have been fully utilized and the demand for transportation continues to rise, the scenario leads to a sharp increase in transportation prices. This situation in the transportation context mirrors the behavior observed in the liquid phase of the VDW state equation for real gases. In the VDW model, once the gas is compressed to a point where it becomes liquid (which represents the end of the plateau phase), any further reduction in volume (or, analogously, any further increase in demand in the transportation context) results in a significant increase in pressure (or prices in the transportation scenario). This is because the liquid phase is much less compressible than the gas phase, just as the transportation system becomes much less flexible once carpooling capacities are maximized.
\end{itemize}

\noindent 
Extending the carpooling analogy to incorporate a multi-agent or microscopic perspective, like for the ideal gas concept, provides a nuanced view of individual decision-making in the context of commuting. In the VDW context, phase transitions between gas and liquid states are influenced by intermolecular attractive forces. Drawing a parallel to carpooling, ecological incentives to share rides can be viewed as analogous to these attractive forces, encouraging individuals to pool together in a single vehicle. Envisioning an "ideal gas" of commuters, where each commuter shares common socio-economic characteristics, allows for an exploration of the individual decision to commute alone or to carpool. This decision-making process involves balancing personal wealth against ecological consciousness. In this analogy, $\kappa_r$, representing the average wealth or economic capacity of commuters, plays a role similar to temperature in the VDW context. At high values of $\kappa_r$ (analogous to high temperatures), individualistic tendencies outweigh ecological considerations, leading to a preference for travelling alone. This scenario aligns with the gas phase in the VDW model, where high temperatures mitigate the effects of attractive forces, preventing the formation of a liquid phase or, in our analogy, a cohesive carpooling group (the VDW plateau). Conversely, for $\kappa_r<1$, mirroring lower temperatures in the VDW framework, the conditions become favorable for the emergence of a \textit{substitution plateau}. This plateau represents a phase where individual and pooled commuting behaviors coexist and interchange. In this extended analogy, the parameter $b$ reflects the limitation on how many commuters can realistically share a single vehicle, which corresponds to the volume exclusion principle in the VDW equation. The parameter $a$ quantifies the strength of the ecological incentives or social pressures that encourage carpooling, analogous to the attractive forces between molecules in the VDW context. 

\subsection{Medical context - generic and original medication}

\noindent Consider the scenario where two treatments, $A$ (an original medication) and $B$ (a generic version), are available for a specific disease. Both treatments contain the same active molecules, but they may have different excipients, with $B$ being more cost-effective. Despite this, patients tend to prefer $A$ due to its established reputation and the perceived added value of its excipients.

Here again, an analogy with the VDW state equation may help understand patient preferences and economic dynamics in healthcare. Let $Y_r$ represent the cost to cure the illness, and $X_r$ the inverse demand for treatment, which inversely correlates with the spread of the illness. The parameter $\kappa_r$ measures the available healthcare budget. For high healthcare budgets ($\kappa_r > 1$), patients predominantly choose the more expensive treatment $A$, analogous to the single gas phase in the VDW state equation where there is little to no compression of particles, representing the minimal economic pressure to switch to a more cost-effective option. As the budget constraint tightens ($\kappa_r < 1$), a portion of the patient population shifts to the generic treatment $B$. This shift creates a substitution equilibrium, similar to the liquid-gas equilibrium in the VDW model, where there is a balance between the original and generic treatments. During this phase, the overall cost of treatment stabilizes, in line with the constant pressure observed on the VDW plateau during a phase transition. Once the entire patient population has switched to the generic option $B$ (analogous to the complete transition to the liquid phase in the VDW model), any further increase in demand (conversely a further decrease in the inverse demand $X_r$) necessitates finding new, potentially costly alternatives. This could involve developing new supply chains or treatments, leading to a sharp increase in $Y_r$. This phase mirrors the incompressibility of the liquid phase in the VDW equation, where further compression (or increased demand in our context) leads to a significant rise in pressure (or cost).

\subsection{Electricity demand  - actual data fitting} \label{ELECTRO} 

\noindent To explore the possibility of applying a state equation similar to Eq.(\ref{REDUCED1X}) (or equivalently  Eq.(\ref{QUADRA2X})) for data fitting purposes, we draw inspiration from study \cite{wan2022}, which focuses on the electricity demand market. 

In their study, the authors successfully employ a cubic polynomial equation to fit actual market data within a specified range. The price-and-demand equation they propose is as follows:

\begin{equation}
\label{CUCU}
Y= a_0 + a_1X + a_2X^{2} -  a_3X^{3} , \qquad a_3 >0
\end{equation}

\noindent Here, $Y$ represents the price, and $X$ represents demand. The coefficients $\left\{a_0, a_1, a_2 ,a_3 \right\}$ are determined through data fitting (see Figure 2 in \cite{wan2022}), effectively capturing the relationship between price and demand in the electricity market. Inspired by this approach, we can consider applying the VDW state equation, Eq.(\ref{QUADRA2X}), to a similar data fitting exercise. Considering the cubic polynomial applied in \cite{wan2022} and the nonlinear properties of the VDW equation, this method could provide a robust framework for modeling the intricate relationships between variables such as price and demand across different markets. The critical step involves meticulously gathering and examining data, followed by adjusting the equation's coefficients to achieve the best fit. Ultimately, this process would lead to insightful interpretations of market dynamics.

\noindent Refer to Figure 2 in \cite{wan2022}, particularly the shaded area. The inverse electricity demand at the center of this shaded area is approximately located at $X \approx 30'000 \,[ {\rm MWh}]$. At this point, the corresponding price is roughly $Y=45  \left[ \frac{{\rm EUR}}{{\rm MWh}}\right]$. Therefore, the slope $\frac{dY}{dX}$ of the straight line (the slope of the blue curve in Figure 2 of \cite{wan2022}) is approximately given by:
\begin{equation}
\label{PPRIME}
Y'(30'000) = \frac{50}{36'000-20'000}= 0.003 \left[ \frac{{\rm EUR}}{{\rm MWh}^{2}}\right]. 
\end{equation}

\noindent From Eq.(\ref{PPRIME}), we observe a very small slope. Within a VDW modeling framework, this strongly indicates that the iso-line described by Eq.(\ref{PPRIME}) is closely aligned with the critical VDW isotherm. Consequently, we can approximately equate the values $Y_c= 45\,[{\rm EUR}][{\rm GWh}] $ and $X_c= 30 [ {\rm GWh}]$ to those at the VDW critical point, facilitating direct model calibration. The corresponding VDW parameters $a$, $b$ and the variable $\kappa_c$ can then be directly identified using Eq.(\ref{REDUCEDX}), namely\footnote{For clarity, note that we use the conversion $1 [ {\rm GWh}] := 1'000 [ {\rm MWh}].$}:
\begin{equation}
\label{CRITOS}
\left\{
\begin{array}{l}
Y_c \simeq  45\,  \left[ {\rm EUR/MWh}\right] = 4.5 \cdot 10^{-2} \, [{\rm  EUR/GWh}]  , \\\\

b= \frac{X_c }{2}= 15 \,[ {\rm GWh}], \\\\ a= 27  Y_c \,b^{2} =3^{7}\cdot 5^{3}  \cdot 10^{-2}=  273, 375 \,\,[{\rm EUR}][{\rm GWh}] , \\\\

\kappa_c = \frac{8a}{27 b} = 5.4 \, [{\rm EUR}].

\end{array}
\right.
\end{equation}

\noindent Based on the parameters $\left\{a,b, \kappa_c \right\}$ obtained from Eq.(\ref{CRITOS}), it is straightforward to formulate a cubic equation model that replicates the identical triple point. The model is expressed as follows:
\begin{equation}
\label{FITOSS}
\left\{
\begin{array}{l}
Y_{{\rm VDW}} = 45 \left[\frac{8}{\frac{X}{15}} - \frac{27}{(\frac{X}{15} +1)^{2}}\right], \qquad  \quad \textcolor{black}{(van\,\, der \,\, Waals\,\, modeling),}\\\\
Y_{{\rm cub}} = -(X-30)^{3} + 45, \qquad \qquad \, \textcolor{black}{(cubic \,\, polynomial \,\, modeling)}.
\end{array}
\right.
\end{equation}

\begin{figure}[htbp]
 \begin{center}
    \includegraphics[height=7cm]{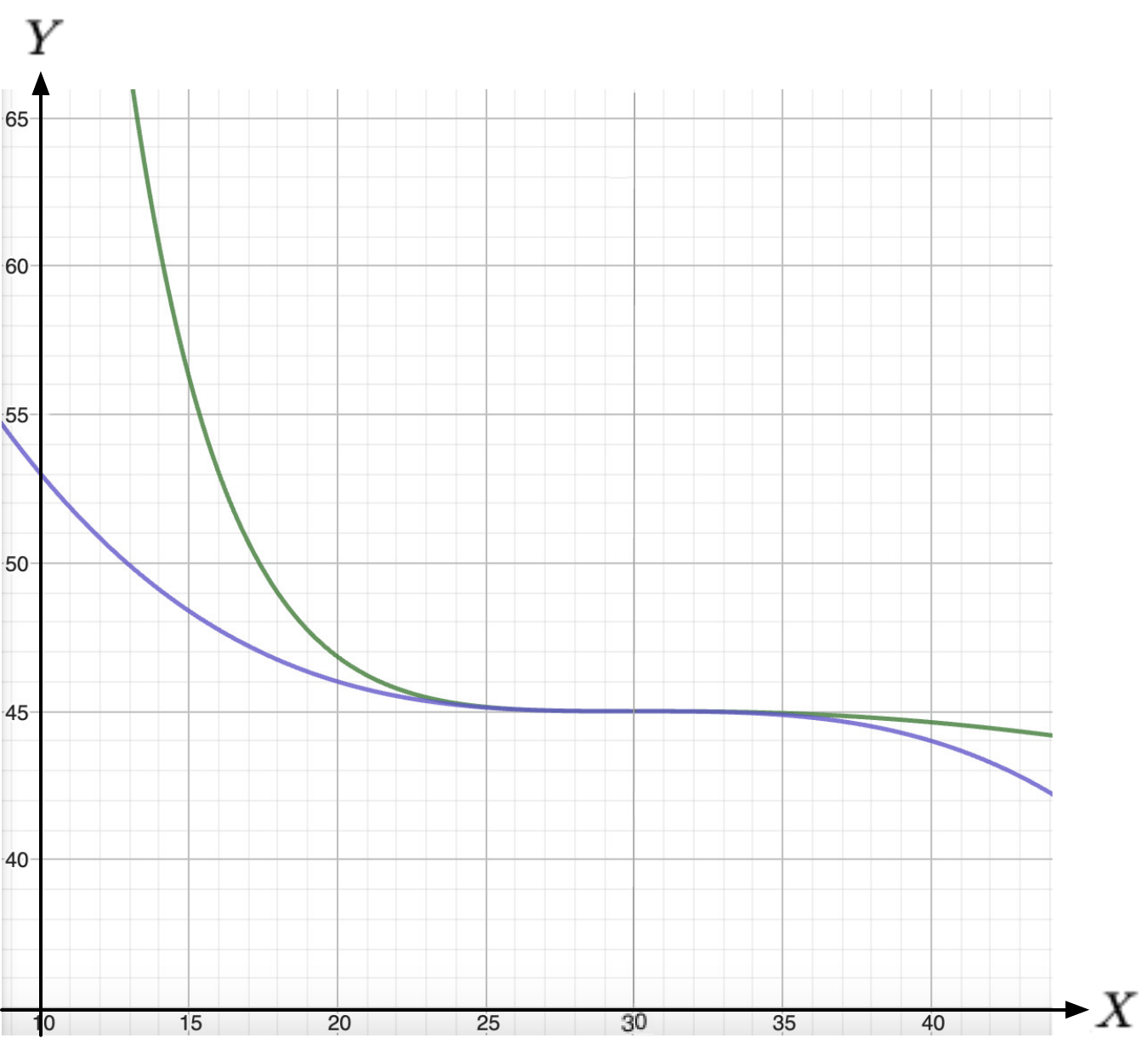}
    \caption{\scriptsize {Fitting of the cubic polynomial (blue) and VDW (green) curves  at the quasi-plateau level $\kappa_c$. From Figure 2 in \cite{wan2022}, one qualitatively estimates that we are closed to a plateau (i.e., close  to the VDW triple point). Assume that we are at the critical level with the triple point  at demand level $X=30 GWh$, as observed in Figure 2 in \cite{wan2022}. We draw the corresponding price-and-demand state equations Eq.(\ref{FITOSS}) with price $Y$ in [EUR/MWh] as ordinate and demand $X$ in [GWh = 1000MWh] as abscissa. There is virtually no usable distinctions between the two curves in the range considered in \cite{wan2022}. However, the VDW curve appears to be steeper away from the plateau, which actually matches the actual monitoring displayed (with black dots) in Figure 2 of \cite{wan2022}.}}
     \label{FITONE}
     \end{center}
 \end{figure}

\noindent The couple of inverse demand curves $Y_{{\rm VDW}}$ and $Y_{{\rm cub}}$  are drawn in Figure \ref{FITONE}.

\vspace{0.3cm}
\noindent {\it Remark}. It is important to note that the cubic polynomial relation in Eq.(\ref{FITOSS}) has a limited range of applicability. This is because it predicts negative prices for large values of inverse demand $X$, which is not realistic. In contrast, the VDW model maintains positive prices for all values of $X$, making it a more robust framework for pricing. Additionally, the elasticity behavior in the VDW model remains coherent across the entire range of inverse demands $X$, unlike the cubic model, which further highlights the VDW model's superior capacity for modeling.

\subsection{Giffen and Veblen behaviors - luxury and inferior goods} \label{VEBLEN}

\begin{figure}[htbp]
 \begin{center}
    \includegraphics[height=6.5cm]{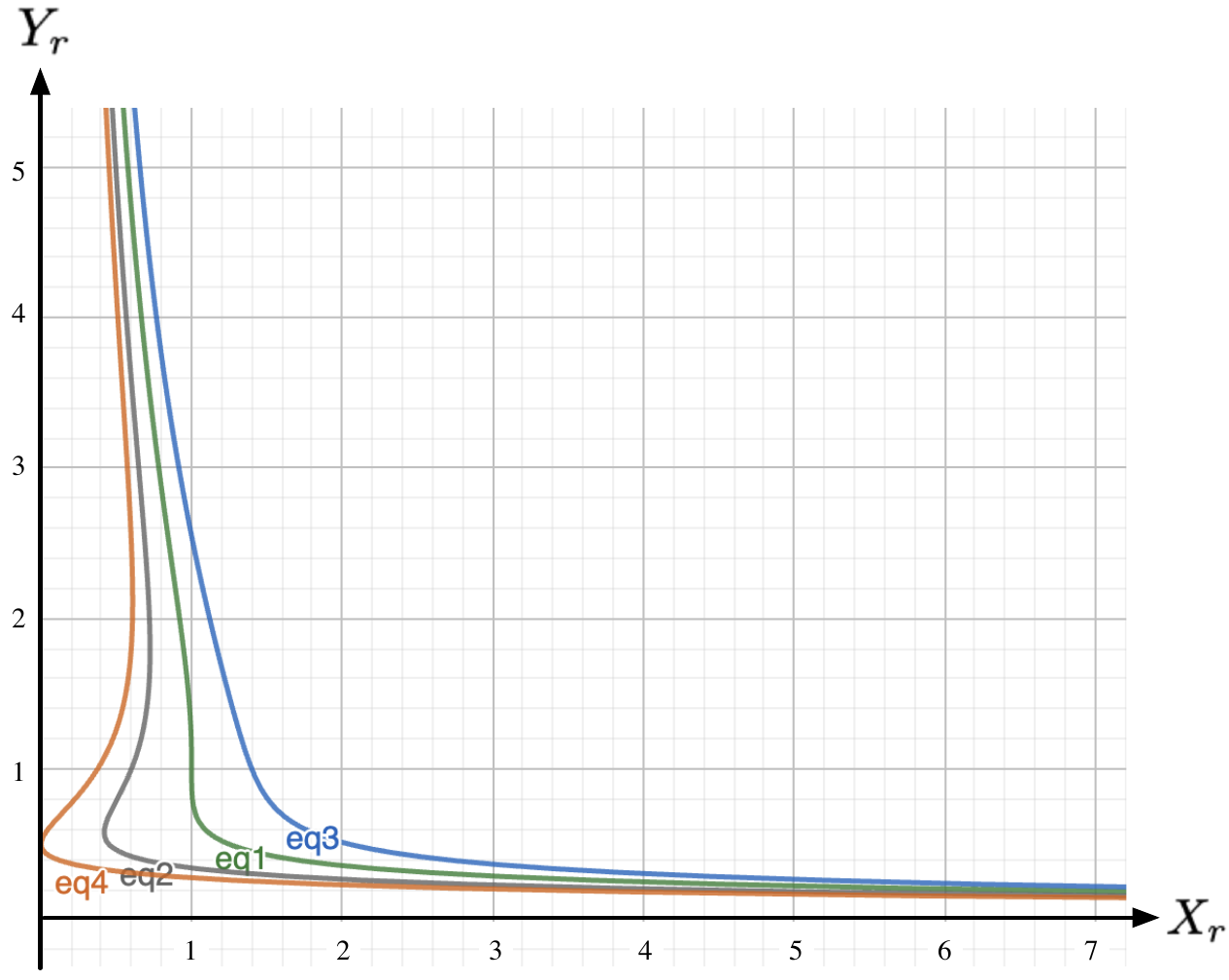}
    \caption{\scriptsize {Iso-budget lines as given by line \textit{(ii)} of Eq.  (\ref{QUADRA2X}) for $\kappa_r = 1.1$ (blue), $\kappa_r = 1.0$ (green), $\kappa_r = 3.6$ (grey), and $\kappa_r = \frac{27}{32}$ (red, corresponding  as before to the $\kappa_r$ lower bound). 
    The  abscissa stands for the  demand $X_r$ and the ordinates for the price $Y_r$. For $\kappa_r <1$, a $Z$-shape behavior emerges. This corresponds to the economically counter-intuitive response  where a reduction of the price induces a reduction of the demand.}}
     \label{GIFFX}
     \end{center}
 \end{figure}

\noindent Let us assume here that the input variable is the price, and with each price  increase, the equilibrium imposes a drop in demand. Accordingly, the adapted modeling  can be described by line \textit{(ii)} of  Eq.  (\ref{QUADRA2X}),  where $Y \rightarrow Y_r$ is the price,  $X \rightarrow X_r$ denotes the demand, and $\kappa \rightarrow \kappa_r$ stands for the income (i.e., the average available budget)\footnote{Here, and contrary to section \ref{SCARCITY}, the variable  $X_r$ is the demand itself and not the inverse demand.}. From Eq.(\ref{QUADRA2X}), a few  corresponding iso-budget lines for a selection of $ \kappa_r$ are derived in Figure \ref{GIFFX}.

 \noindent With line \textit{(ii)} of Eq.(\ref{QUADRA2X}), the corresponding elasticity ${\cal E}_{Y_r/X_r}$ is given by  Eq.(\ref{REDUCED2X}), with a $X_r \leftrightarrow Y_r$ exchange of the variables, and we have:

 \begin{equation}
\label{EXCHANGE}
 {\cal E}_{Y_r/X_r} :=  \frac{\left\{d \ln Y_r\right\}}{\left\{d \ln X_r\right\}} =  -  \left[ 
 \frac
 {\frac{4\kappa_r}{X_r} - 27[Z(X_r) ]^{2}}
 {
 \frac{4\kappa_r}{X_r} - 108 [Z(X_r) ]^{3}}
\right].
 \end{equation}
 
 \noindent We again observe a singularity at  $\kappa_r =1 \Rightarrow {\cal E}_{Y_r/X_r}= \infty$. For $\kappa_r >1$, the resulting elasticity $ {\cal E}_{Y_r/X_r} $ is systematically  negative, thus describing a standard behavior. For $\kappa_r <1$ however, the iso-budget lines exhibit $Z$-shaped curves. As in Section \ref{SCARCITY}, a (here vertical) plateau can also be constructed. However, it is worth  pointing out that, in this case, the $Z$-shape itself offers a natural modeling framework for the so-called Veblen  and/or Giffen behaviors (see among others \cite{dougan1982} and \cite{philbois2018}). Veblen/Giffen effects are induced by the absence of economic substitution alternatives. Hence, the  full $Z$-shape behavior is here needed instead of a VDW-type plateau. The Veblen effect is observed in luxury markets, where the prestige associated with luxury goods means that a decrease in price can lead to reduced demand, as the goods lose some of their prestige value. On the other hand, the Giffen effect pertains to certain inferior goods, such as staple foods like rice and bread. In this scenario, a price increase paradoxically boosts demand because, for consumers with limited budgets, the alternatives become prohibitively expensive. For example, a typical consumer might balance their daily food intake between bread and meat. With a reduced budget (i.e.\,$\kappa_r<1$), they may no longer be able to afford meat, compelling them to consume more bread to meet their dietary needs.
 \section{Conclusion and perspectives}\label{conclusion}
 
\noindent 

The Van der Waals (VDW) state equilibrium equation, with its minimal reliance on exogenous parameters, offers a comprehensive and unifying framework for analyzing both substitution effects in economics and the peculiar phenomena of Giffen and Veblen behaviors. In general, the VDW model, with its capability to accommodate non-constant elasticity curves, emerges as a compelling and natural alternative to the commonly used cubic and exponential models for quantitative fitting. Furthermore, the VDW theory, characterized by its depiction of a cusp catastrophe singularity, is well-suited for capturing market equilibria involving key variables such as price, demand, and budget. Drawing on approaches from physics, VDW modeling bridges micro- and macro-economic models and perspectives through the possibility of a mean-field description of multi-agent systems. In this context, the VDW inter-molecular attractive forces manifest a propensity among  agents to adopt common consensual behaviors, as seen for example in carpooling practices. With over 150 years of successes in thermostatics, the VDW state equation may now be poised for a new  cross-disciplinary \textit{life}, a potential that was already highlighted by the late Fran\c{c}ois Roddier, \cite{roddier2017}.

\bibliographystyle{abbrvnat} 
\bibliography{references} 

\begin{thebibliography}{11}
\providecommand{\natexlab}[1]{#1}
\providecommand{\url}[1]{\texttt{#1}}
\expandafter\ifx\csname urlstyle\endcsname\relax
  \providecommand{\doi}[1]{doi: #1}\else
  \providecommand{\doi}{doi: \begingroup \urlstyle{rm}\Url}\fi

\bibitem[Dougan(1982)]{dougan1982}
W.~R. Dougan.
\newblock Giffen goods and the law of demand.
\newblock \emph{Journal of Political Economy}, 90\penalty0 (4):\penalty0
  809--815, 1982.

\bibitem[Guimaraes and Sheedy(2011)]{guimaraes2011}
B.~Guimaraes and K.~D. Sheedy.
\newblock Sales and monetary policy.
\newblock \emph{American Economic Review}, 101\penalty0 (2):\penalty0 844--876,
  2011.

\bibitem[Gumjudpai(2018)]{gumjudpai2018}
B.~Gumjudpai.
\newblock Towards equation of state for a market: A thermodynamical paradigm of
  economics.
\newblock \emph{Journal of Physics: Conference Series, Siam Physics Congress
  2018}, 1144:\penalty0 10.1088/1742--6596/1144/1/012181, 2018.

\bibitem[Huang et~al.(2014)Huang, Liu, Zhoua, Zhang, and Wang]{huang2014}
Q.~Huang, Z.~Liu, Y.~Zhoua, D.~Zhang, and F.~Wang.
\newblock Study on mechanisms of co2 bleve based on the cusp- catastrophe
  model.
\newblock \emph{Energy Procedia}, 61:\penalty0 1343--1347, 2014.

\bibitem[Jammernegg and Fischer(1986)]{jammernegg1986}
W.~Jammernegg and E.~O. Fischer.
\newblock Economic applications and statistical analysis of the cusp
  catastrophe model.
\newblock \emph{Zeitschrift fuer Operations Research}, 30:\penalty0 B45--B58,
  1986.

\bibitem[Levine(1978)]{levine1978}
I.~Levine.
\newblock \emph{Physical Chemistry}.
\newblock McGraw-Hill, 1978.

\bibitem[Philbois and Block(2018)]{philbois2018}
G.~Philbois and W.~E. Block.
\newblock The z curve: Supply and demand for giffen goods.
\newblock \emph{MISES: Interdisciplinary Journal of Philosophy Law and
  Economics}, 6\penalty0 (3):\penalty0 503--508, 2018.

\bibitem[Rashkovskiy(2017)]{rashkovskiy2021}
S.~Rashkovskiy.
\newblock Economic thermodynamics.
\newblock \emph{Physica A: Statistical Mechanics and its Applications},
  582:\penalty0 126261, 2017.

\bibitem[Roddier(2017)]{roddier2017}
F.~Roddier.
\newblock L'\'equation de van der waals appliqu\'ee \`a l'\'economie.
\newblock \emph{Res-Systematica}, 16:\penalty0 02, 2017.

\bibitem[Saslow(1999)]{saslow1999}
W.~M. Saslow.
\newblock An economic analogy to thermodynamics.
\newblock \emph{American Journal of Physics}, 67:\penalty0 1239--1247, 1999.

\bibitem[Wan et~al.(2022)Wan, Kober, and Densing]{wan2022}
Y.~Wan, T.~Kober, and M.~Densing.
\newblock Nonlinear inverse demand curves in electricity market modeling.
\newblock \emph{Energy Economics}, 107:\penalty0 105809, 2022.

\end{thebibliography}
  
\section*{Appendix}

\noindent In this section, we describe additional analytical properties that follow from  Eq.(\ref{REDUCED1X}).

\vspace{0.2cm}
\noindent {\bf Corollary 1}.
{\it 
\begin{itemize}
  \item[\textit{(i)}]  \quad For  $\kappa_r >1$, we have  $\Delta <0$, and ${\cal R}(X_r,Y_r)= \kappa_r$  defines  a \textbf{one-to-one relation } leading to  ${\cal E}_{Y_r/X_r}<0$.
 
  \item[\textit{(ii)}]  \quad For $\kappa_r >1$, we have $\Delta >0 $, and ${\cal R}(X_r,Y_r)= \kappa_r$  defines  a \textbf{surjective relation} allowing  ${\cal E}_{Y_r/X_r}<0$ within a region $ {\cal W} (\kappa_r):= \left]X_{r}^{g}(\kappa_r), X_{r}^{d}(\kappa_r)\right[$, where $X_{r}^{g}(\kappa_r) \leq X_{r}^{d}(\kappa_r)$ are the respective abscissae of the  $Y_r$-extrema.
\end{itemize}}

\vspace{0.3cm}

\noindent {\bf Proof of Corollary 1}.

\noindent This corollary is an immediate consequence of Proposition 2.

\rightline{$\Box$}

\noindent {\bf Lemma 2}: {\it (miscellaneous properties of the elasticities ${\cal E}_{Y_r/X_r} $ and ${\cal E}_{X_r/\kappa_r} $).}

\begin{itemize}
  \item[\textit{(i)}]  \quad 
  $
\displaystyle \lim_{X_r \rightarrow 0^{+}} {\cal E}_{Y_r/X_r}  =   \lim_{X_r \rightarrow \infty } {\cal E}_{Y_r/X_r}  =-1.
$
\vspace{0.2cm}
  \item[\textit{(ii)}] \quad {\it ${\cal E}_{ Y_r/X_r=\frac{1}{2}} = -1$ for all $\kappa_r$}.
  
\vspace{0.2cm}
  \item[\textit{(iii)}]  \quad $
   \kappa_r > \frac{27}{32}\quad  \Rightarrow \quad {\cal E}_{Y_r/X_r}=
 \left\{
\begin{array}{l}
< 0 \qquad\,\,\,\;  {\rm for } \qquad X_r < \frac{1}{2} , \\\\
-1  \qquad  \quad \,{\rm for} \qquad X_r = \frac{1}{2}.
\end{array}
\right.
$

\vspace{0.2cm}
\item[\textit{(iv)}]  \quad $\kappa_r >1 \quad \Rightarrow \quad {\cal E}_{Y_r/X_r}<0$.

\vspace{0.2cm}
\item[\textit{(v)}] \,\,\, elasticity at the triple point:
${\cal E}_{Y_r=1/ X_r=1}= \infty.$

\vspace{0.2cm}
\item[\textit{(vi)}]  \,\quad for $Y_r > 1\quad  \Rightarrow \quad {\cal E}_{X_r/\kappa_r} >0 $. 

\vspace{0.2cm}
\item[\textit{(vii)}]  \quad for all  $Y_r \quad \Rightarrow \quad \displaystyle \lim_{X_r \rightarrow +\infty } {\cal E}_{X_r/\kappa_r}=\displaystyle \lim_{X_r \rightarrow 0^{+}} {\cal E}_{X_r/ \kappa_r}=1$.

\end{itemize}

%

\vspace{0.3cm}

%
%
%
%

%
%
%
%
%
%
%
%
%



\end{document}